\documentstyle[12pt]{article}

\title{\bf One-loop quantum cosmological correction to the gravitational constant in the closed Friedmann-Robertson-Walker universe}
\author{S. Jalalzadeh$^1$\thanks{email: s-jalalzadeh@sbu.ac.ir} and F. Darabi\thanks{e-mail:
f.darabi@azaruniv.edu}\\{\small Department of
Physics, Shahid Beheshti University, Evin, Tehran 19839,
Iran}\\{\small Department of Physics, Azarbaijan University of
Tarbiat Moallem, 53714-161, Tabriz, Iran .} }
\begin{document}
\maketitle
\begin{abstract}
In this paper, we calculate the one-loop quantum cosmological corrections to the kink energy in the closed Friedmann-Robertson-Walker universe in which the fluctuation potential $V^{\prime\prime}$ has a shape invariance property. We use the generalized zeta function regularization method to implement our
setup for describing quantum kink-like states. It is conjectured that the corrections lead to the renormalized gravitational constant.
\end{abstract}

\newpage
\section{Introduction}

The quantum corrections to the mass of classical topological defects
play an important role in the semi-classical approach to quantum
field theory \cite{1,1'}. The computation of quantum energies around
classical configurations in (1 + 1)-dimensional kinks has been
developed in \cite{2,2',2'',2'''} by using topological boundary conditions, the
derivative expansion method \cite{3,3',3'',3'''',3'''}, the scattering phase shift
technique \cite{4}, the mode regularization approach \cite{5}, the
zeta- function regularization technique \cite{6}, and also the
dimensional regularization method \cite{7}. In a previous paper by
one of the authors, the one-loop renormalized kink quantum mass
correction in a (1 + 1)-dimensional scalar field theory model was
derived using the generalized zeta function method for those
potentials where the fluctuation potential $V^{''}$ has the shape
invariance property \cite{8}. These potentials are very important
since they possess a shape invariant operator in their prefactor
which makes the corrections of this kind of potential exact by the
heat kernel method. This kind of potential occurs in different
fields of physics particularly in quantum gravity and cosmology. For
example, the tunneling rate with exact prefactor was calculated, to
first order in $\hbar$, for an empty closed
Friedmann-Robertson-Walker (FRW) universe with decaying cosmological
term and representing this kind of potential \cite{8'}.

In this paper, we implement the above technics to compute the
quantum corrections to the ground state energy of the Kink-like solutions
of an empty closed Friedmann-Robertson-Walker universe with a
decaying cosmological term. To this end, we shall study the
corresponding minisuperspace and construct the Euclidean action.

\section{Mini-superspace of closed Friedmann-Robertson-Walker universe}

We shall consider an empty closed ($k=1$) Friedmann-Robertson-Walker
universe (FRW) universe with a non vanishing cosmological term
$\Lambda$. The line element is given by
\begin{eqnarray}\label{b}
ds^2=-c^2dt^2+a(t)^2\left[\frac{dr^2}{1-r^2}+r^2d\Omega^2\right],
\end{eqnarray}
where $a(t)$ as the scale factor is the only dynamical degree of
freedom. The pure gravitational action corresponding to (\ref{b}) is given
by
$$
S=\frac{c^4}{16\pi G}\int_{\mathcal
M}d^4x\sqrt{-g}(R-2\Lambda)+\frac{c^2}{8\pi G}\int_{\partial{\mathcal
M}}d^3x\sqrt{g^{(3)}}K
$$
$$
=\frac{3c^2\pi}{4G}\int dt \left[-a\dot{a}^2+c^2a-\frac{\Lambda}{3}a^3\right]:=
\int dt L.
$$
\begin{equation}\label{2}
=\frac{m}{2}\int dt \left[-a
{\dot{a}}^2+a\left(c^2-\frac{\Lambda}{3} a^2 \right)
\right], \label{d}
\end{equation}
where  $m=3\pi c^2/(2G)$.
Assuming a positive cosmological constant $\Lambda$ leads to an exponentially
expanding universe so called $de\: Sitter$ universe. On the other hand,
defining the cosmological constant as the constant vacuum energy
density $\rho_{vac}$, coming from condensation of scalar fields, results
in Guth's inflationary model. From quantum field theory calculations we 
expect $\rho_{vac}\sim M_p^4$ which leads to an extraordinarily rapid
expansion of the scale factor. However, this value of cosmological
constant is enormous when compared with observations. This is the
cosmological constant problem. One possible solution to this problem
is to take a decaying cosmological constant \cite{SW}
\begin{equation}
\Lambda(a)=\Lambda(a_0) \left(\frac{a_0}{a}\right)^{m} \label{f}
\end{equation}
where $a_0$ is the value of the scale factor at an arbitrary
reference time and $m$ is an arbitrary constant. Let us now assume $m=2-\frac{2}{\alpha}$
to cast the cosmological term in its appropriate form as
\begin{equation}
\Lambda(a)=\Lambda(a_0)
\left(\frac{a_0}{a}\right)^{2-\frac{2}{\alpha}}, \label{5}
\end{equation}
where $\alpha>0$ is taken for convenience. Substituting (\ref{5})
into the action (\ref{d}) leads to
\begin{equation}
I=\frac{m}{2}\int \! dt \left[-a
{\dot{a}}^2+a\left(c^2-\left(\frac{a}{a_0}\right)
^{\frac{2}{\alpha}} \right) \right], \label{g}
\end{equation}
where $\Lambda(a_0)=\frac{3}{{a_0}^2}$. The Euclidean form of the
action (\ref{g}) is not suitable to be used in instanton calculation
techniques. The reason is that the kinetic term is not in its
standard quadratic form. It has been recently shown that in such
cosmological model one may use the Duru-Kleinert equivalence to work
with the standard form of the action \cite{13}, \cite{12}. Using the
same procedure, we find the Duru-Kleinert equivalent action in the
present cosmological model as follows
\begin{equation}
I_0=\frac{m}{2}\int\! dt \left[
{\dot{a}(\tau)}^2+a^2\left(c^2-\left(\frac{a}{a_0}\right)
^{\frac{2}{\alpha}} \right) \right] \label{h}
\end{equation}
Now, the Euclidean action (\ref{h}) has the right kinetic term to be
used in the instanton calculations. The Euclidean type Hamiltonian
corresponding to the action (\ref{h}) is given by
\begin{equation}
H_{E} = \frac{m}{2}\dot{a}^2 -\frac{m}{2}a^2
\left[c^2-\left(\frac{a}{a_0}\right) ^{\frac{2}{\alpha}} \right]
\end{equation}
whose vanishing constraint $H_{E}=0$ \footnote{The constraint
$H_{E}=0$ corresponds to Euclidean form of the Einstein equation.}
gives a non-trivial instanton (kink) solution
\begin{equation}
a(\tau)= \frac{a_0}{(\cosh(\frac{\tau}{\alpha}))^\alpha} \label{i}
\end{equation}
corresponding to the potential
\begin{equation}
V(a) = \frac{m}{2}a^2 \left[c^2-\left(\frac{a}{a_0}\right)
^{\frac{2}{\alpha}} \right]\:\:\:\:\:For \:a\geq0. \label{j}
\end{equation}
Each solution with $\alpha>0$ describes a particle rolling down from
the top of the potential $ -V(a) $ at $ \tau \rightarrow -\infty $
and $ a = 0 $, bouncing back at $ \tau = 0 $ and $ a = a_0 $ and
finally reaching the top of the potential at $ \tau \rightarrow
+\infty $ and $ a = 0 $. In the next section, we quote briefly how
can we calculate the zeta function of an operator through the heat
kernel method.

\section{Semi-Classical Soliton States}

Classical configuration space is found by static configuration
$\Phi(x)$, so that the energy functional
\begin{eqnarray}\label{21}
E[\Phi]=\int dx\left[\frac{1}{2}\Phi_{,\mu}
\Phi^{,\mu}+V(\Phi)\right],
\end{eqnarray}
is finite. One can describe quantum evolution in  Schrodinger
picture by the following functional equation
\begin{eqnarray}\label{22}
i\hbar\frac{\partial}{\partial t}\Phi[\phi(x),t]=H\Phi[\phi(x),t],
\end{eqnarray}
so that  quantum Hamiltonian operator is given by
\begin{eqnarray}\label{23}
H=\int
dx\left[-\frac{\hbar^2}{2}\frac{\delta}{\delta\phi(x)}\frac{\delta}{\delta\phi(x
)}+E[\phi]\right].
\end{eqnarray}
In the field representation the matrix elements of evolution operator are
given by
\begin{eqnarray}\label{24}
\begin{array}{cc}
G(\phi^{(f)}(x),\phi^{(i)}(x),T)=\langle\phi^{(f)}|e^{-\frac{iT}{\hbar}H}|\phi^{
(i)}\rangle\\
\\
=\int D[\phi(x,t)]\exp{(\frac{-i}{\hbar}S[\phi])},
\end{array}
\end{eqnarray}
where the initial conditions are those of static kink solutions of
classical equations where $\phi^{(i)}(x,0)=\phi_{k}(x)$,
$\phi^{(f)}(x,T)=\phi_{k}(x)$. In  semi-classical picture, we are
interested in loop expansion for evolution operator up to the  first
quantum correction
\begin{eqnarray}\label{25}
\begin{array}{cc}
G(\phi^{(f)}(x),\phi^{(i)}(x),\beta)=\\
\\
\exp{(-\frac{\beta}{\hbar}E[\phi_k])}
Det^{-\frac{1}{2}}\left[-\partial^2_{\tau}+P\Delta\right](1+{\cal
O}(\hbar)),
\end{array}
\end{eqnarray}
where we use analytic continuation to Euclidean time,
$t=-i\tau$,$T=-i\beta$, and $\Delta$ is the differential operator
\begin{eqnarray}\label{26}
 \Delta=-\frac{d^2}{d
 x^2}+\frac{d^2V}{d\phi^2}\mid_{\phi=\phi_k},
\end{eqnarray}
P is the projector over the strictly positive sector of the spectrum of
$\Delta$
\begin{eqnarray}\label{27}
\Delta\xi_n(x)=\omega_n^2\xi_n(x),\,\,\,\,\omega^{2}_{n}\,\,
\epsilon \,\, Spec(\Delta)= Spec(P\Delta)+\{0\}.
\end{eqnarray}
We write the functional determinant in the following form
\begin{eqnarray}\label{28}
Det\left[-\frac{\partial^2}{\partial\tau^2}+\Delta\right]=\prod_{n}det\left[-\frac{\partial^2}{
\partial\tau^2}+\omega^{2}_{n}\right].
\end{eqnarray}
All determinants in the infinite product correspond to harmonic
oscillators of frequency $\omega_n$. On the other hand, it is well
known that \cite{10}
\begin{eqnarray}\label{29}
\begin{array}{cc}
det\left(-\frac{\partial^{2}}{\partial \tau^2}+\omega^2_n
\right)^{-\frac{1}{2}}
= \prod_{j=1}^N \left(\frac{j^2\pi^2}{\beta^2}+\omega_n^2\right)^{-\frac{1}{2}}\\
\\
=
\prod_j\left(\frac{j^2\pi^2}{\beta^2}\right)^{-\frac{1}{2}}\prod_j\left(1+\frac{\omega^2_n\beta^2}{j^2\pi^2}\right)^{-\frac{1}{2}}.
\end{array}
\end{eqnarray}
The first product dose not depend on $\omega_n$ and combines with
the Jacobian and other factors we have collected into a single
constant. The second factor has the limit
$\left[\frac{\sinh(\omega_n\beta)}{\omega_n\beta}\right]^{-\frac{1}{2}}$. Therefore, for large $\beta$ with an appropriate normalization we obtain \begin{eqnarray}\label{30}
\begin{array}{cc}
G(\phi^{(f)}(x),\phi^{(i)}(x),\beta)\cong\\
\\
\exp{(-\frac{\beta}{\hbar}E[\phi_k])}\prod_{n}(\frac{\omega_n}{\pi\hbar})^{\frac{1
}{2}}\exp{\left(-\frac{\beta}{2}\sum_{n}\omega_n(1+{\cal
O}(\hbar))\right)}
\end{array}
\end{eqnarray}
where the eigenvalues in the kernel of $\Delta$ have been excluded.
Interesting eigenenergy wave functionals
\begin{eqnarray}\label{31}
H\Phi_j[\phi_k(x)]=\varepsilon_j\Phi_j[\phi_{k}(x)]
\end{eqnarray}
we have an alternative expression for $G_E$ as 
$\beta\rightarrow\infty$
\begin{eqnarray}\label{32}
\begin{array}{cc}
G(\phi^{(f)}(x),\phi^{(i)}(x),\beta)\cong\\
\\
\Phi^{*}_{0}[\phi_k(x)]\Phi_{0}[\phi_k(x)]\exp{(-\beta\frac{\varepsilon_0}{\hbar
})},
\end{array}
\end{eqnarray}
and therefore, using (\ref{30}) and (\ref{32}), we obtain
\begin{eqnarray}\label{33}
\varepsilon_0=E[\phi_k]+\frac{\hbar}{2}\sum_{\omega^{2}_{n}>0}\omega_n+{\cal
O}(\hbar^2),
\end{eqnarray}
\begin{eqnarray}\label{34}
|\Phi_0[\phi_k(x)]|^2=Det^{\frac{1}{4}}\left[\frac{P\Delta}{\pi^2\hbar^2}\right],
\end{eqnarray}
as the Kink ground state energy and wave functional up to One-Loop
order.\\
If we define the generalized zeta function
\begin{eqnarray}\label{35}
\zeta_{P\triangle}(s)=Tr(P\Delta)^{-s}=\sum_{\omega^2_n>0}\frac{1}{(\omega^2_n)^
s},
\end{eqnarray}
associated to  differential operator $P\triangle$, then
\begin{eqnarray}\label{36}
\begin{array}{cc}
\varepsilon_0^k=E[\phi_k]+\frac{\hbar}{2}Tr(P\Delta)^{\frac{1}{2}}+{\cal
O }(\hbar^2)
=\\
\\E[\phi_k]+\frac{\hbar}{2}\zeta_{P\Delta}(-\frac{1}{2})+{\cal
O} (\hbar^2).
\end{array}
\end{eqnarray}
The eigenfunction of $\Delta$ is a basis for  quantum fluctuations
around kink background, therefore the sum of associated zero-point
energies encoded in $\zeta_{P\Delta}(-\frac{1}{2})$ is
infinite. According to the zeta function regularization procedure, and
energy-mass renormalization prescription, the renormalized kink
energy in semi-classical limit becomes \cite{11}
\begin{eqnarray}\label{37}
\begin{array}{cc}
\varepsilon^k(s)=E[\phi_k]+\Delta M_k +{\cal O}(\hbar^2)
=E[\phi_k]+\\
\\
\lim_{s\rightarrow\frac{-1}{2}}[\delta_1\varepsilon^k(s)+\delta_
2^k\varepsilon(s)]+{\cal O}(\hbar^2),
\end{array}
\end{eqnarray}
where
\begin{eqnarray}\label{38}
\begin{array}{cc}
\delta_1\varepsilon^k(s)= \frac{\hbar}{2}\mu^{2s+1}[\zeta_{P\Delta}(s)-\zeta_{\nu}(s)],\\
\\
\delta_2\varepsilon^k(s)=
\lim_{L\rightarrow\infty}\frac{\hbar}{2L}\mu^{2s+1}\frac{\Gamma(s+1)}{\Gamma(s)
}
\zeta_\nu(s+1)\times\\
\\
\int_{-\frac{L}{2}}^{\frac{L}{2}}
dx\left[\frac{d^2V}{d\phi^2}|_{\phi_k}-\frac{d^2V}{d\phi^2}|_{\phi\nu}\right].
\end{array}
\end{eqnarray}
Here $\phi_\nu$ is a constant minimum of the potential $V(\phi)$,
$E[\phi_k]$ is the corresponding classical energy, $\mu$ has
the dimension of $length^{-1}$ introduced to make the terms in
$\delta_1\varepsilon^k(s)$ and $\delta_2\varepsilon^k(s)$
homogeneous from a dimensional point of view, and $\zeta_\nu$
denotes the zeta function associated with vacuum $\phi_v$.\\
Now we explain very briefly how can we calculate the zeta function of
an operator through the heat kernel method. We introduce  generalized
Riemann zeta function of operator A by
\begin{eqnarray}\label{39}
\zeta_{A}(s)=\sum_{n}\frac{1}{|\lambda_n|^s},
\end{eqnarray} where
$\lambda_n$ are the eigenvalues of the operator $A$.
  On the other hand, $\zeta_A(s)$ is
 the Mellin transformation of heat kernel $G(x,y,t)$ which satisfies
 the following heat diffusion equation
 \begin{eqnarray}\label{40}
 A G(x,y,t)=-\frac{\partial}{\partial t}G(x,y,t),
 \end{eqnarray}
 with an initial condition $G(x,y,0)=\delta(x-y)$. Note that
 $G(x,y,t)$ can be written in terms of its spectrum
 \begin{eqnarray}\label{41}
 G(x,y,t)=\sum_{n}e^{-\lambda_n t}\psi_n^{*}(x)\psi_n(y),
 \end{eqnarray}
 and as usual, if the spectrum is continues, one should integrate it. From relation (\ref{39}), (\ref{40}) and (\ref{41}) it is clear that
\begin{eqnarray}\label{42}
 \zeta_A(s)=\frac{1}{\Gamma(s)}\int_{0}^{\infty}d\tau\tau^{s-1}\int_{-\infty}^{
 \infty}G(x,x,\tau)dx.
 \end{eqnarray}
 Hence, if we know the associated Green function of an operator,
  we can calculate the generalized zeta function corresponding
 to that operator.

\section{Renormalized ground state energy of the
Kink-like solution of closed Friedmann-Robertson-Walker universe}

Comparison of the sections two and three reveals that
the system of closed Friedmann-Robertson-Walker universe may be equivalent
to the classical configuration space of the static field
$\Phi(x)$ from the shape invariance point of view. Therefore, the one-loop quantum corrections to the mass quantity of closed Friedmann-Robertson-Walker universe, namely $m=3\pi c^2/(2G)$, is equivalent to compute the one-loop quantum corrections to the kink mass or ground state energy of the scalar field $\Phi(x)$. We call this procedure as the one-loop quantum corrections to the ground state energy of the Kink-like solution of closed Friedmann-Robertson-Walker universe. In other words, what is physically meant by the one-loop quantum corrections to the ground state energy in FRW universe is nothing but the
one-loop quantum corrections to the mass quantity $m=3\pi c^2/(2G)$ in FRW
universe which resembles the kink's mass or energy in the equivalent system
of the classical static field $\Phi(x)$ configuration.  

In this regard, we need the spectrum of differential operator (\ref{26}) and the corresponding
vacuum. We assume
$\alpha$ to be positive integer. According to the previous
section, the operator (\ref{26}) which acts on the eigenfunctions becomes\footnote{The combination $\frac{mc^2}{\hbar}$
arises naturally due to the well known
fact that in every quantum mechanical problem of gravity $\hbar$ appears
together with $m$ \cite{Sakurai}.}
\begin{eqnarray}\label{44}
\Delta_{l+h} =\frac{mc^2}{\hbar}\left(-\frac{d^2}{dx^2} +l^2-\frac{l(l+1)}{\cosh^2 x}+h\right),
\end{eqnarray}
where $x=\sqrt{m}c\tau$, $l=\alpha+1$ and $h=1-2l=-(1+2\alpha)$.   Also the operator acting on the vacuum has the following form
\begin{eqnarray}\label{45}
\Delta_{l+h}(0)=\frac{mc^2}{\hbar}\left(-\frac{d^2}{dx^2}+l^2+h\right).
\end{eqnarray}
Note that we have the constant shift $h$ in the spectrum that we add it on
our calculations latter (see Eq.({\ref{71})). 
Also, since $\zeta_{\sigma\Delta}(s)=|\sigma|^{-s}\zeta_\Delta(s)$
then we may ignore $\sigma=\frac{mc^2}{\hbar}$ here in  our calculations and introduce it again at last steps.

In the reminding of this section, to obtain the spectrum of
(\ref{44}) we will use the shape invariance property. First we
review briefly the important concepts that we will use.
Consider the following one-dimensional bound-state Hamiltonian
\begin{eqnarray}\label{46}
H= -\frac{d^2}{dx^2}+U(x), \hspace{2cm} x\in I \subset {\cal R},
\end{eqnarray}
where $I$ is the domain of  $x$ and $U(x)$ is a real function of
$x$, which can be singular only in the boundary points of the
domain. Let us denote the eigenvalues and
eigenfunctions of $H$ by $E_n$ and $\psi_n(x)$, respectively. We use factorization method
which consists of writing Hamiltonian as the product of two first
order mutually adjoint differential operators $A$ and $A^\dagger$.
If the ground state eigenvalue and eigenfunction are known, then
one can factorize Hamiltonian (\ref{46}) as
\begin{eqnarray}\label{47}
H=A^\dagger A +E_0,
\end{eqnarray}
where $E_0$ denotes the ground-state eigenvalue,
\begin{eqnarray}\label{48}
\begin{array}{lll}
A=\frac{d}{dx}+W(x),\\
\\
A^\dagger=-\frac{d}{dx}+W(x),
\end{array}\
\end{eqnarray}
and
\begin{eqnarray}\label{49}
W(x)=-\frac{d}{dx}\ln(\psi_0).
\end{eqnarray}
Supersymmetric quantum mechanics (SUSY QM) begins with a set of two
matrix operators, known as supercharges
\begin{equation}
Q^+ =\left(\begin{array}{cc} 0 & A^\dagger\\
\\0 & 0 \end{array}\right)_, \hspace{2cm} Q^- =\left(\begin{array}{cc} 0 & 0\\
\\A & 0 \end{array}\right)_.\label{50}
\end{equation}
This operators form the following superalgebra \cite{14}
\begin{eqnarray}\label{51}
\{Q^+,Q^-\}=H_{SS}, \hspace{1cm} [H_{SS},Q^{\pm}]=(Q^\pm)^2=0,
\end{eqnarray}
where SUSY Hamiltonian $H_{SS}$ is defined as
\begin{equation}
H_{SS}=\left(\begin{array}{cc} A^\dagger A & 0\\
\\0 & AA^\dagger \end{array}\right)=\left(\begin{array}{cc} H_1 & 0\\
\\0 & H_2 \end{array}\right)_.\label{52}
\end{equation}
In terms of the Hamiltonian and supercharges
\begin{eqnarray}\label{53}
\begin{array}{ccc}
Q_1=\frac{1}{\sqrt{2}}(Q^++Q^-),\\
\\
Q_2=\frac{1}{\sqrt{2i}}(Q^+-Q^-),
\end{array}
\end{eqnarray}
the superalgebra takes the form
\begin{eqnarray}\label{54}
\{Q_i,Q_j\}=H_{SS}\delta_{ij}, \hspace{0.5cm}[H_{SS},Q_i]=0,
\hspace{0.5cm} i,j=1,2.
\end{eqnarray}
The operators $H_1$ and $H_2$ defined by
\begin{eqnarray}\label{55}
\begin{array}{cc}
H_1= A^\dagger  A = -\frac{d^2}{dx^2} +U_1=-\frac{d^2}{dx^2}+W^2-\frac{dW}{dx},\\
\\
H_2=AA^\dagger = -\frac{d^2}{dx^2} + U_2
=-\frac{d^2}{dx^2}+W^2+\frac{dW}{dx},
\end{array}
\end{eqnarray}
are called SUSY partner Hamiltonians and the function $W$ is called
the superpotential. Now, let us denote the eigenfunctions of $H_1$ and $H_2$
with eigenvalues $E^{(1)}_{l}$ and $E^{(2)}_l$ by $\psi^{(1)}_{\,\,\, l}$
and $\psi^{(2)}_{\,\,\, l}$, respectively. It is
easy to see that the eigenvalues of the above Hamiltonians are
positive and isospectral, i.e., they have almost the same energy
eigenvalues, except for the ground state energy of $H_1$. According
to \cite{14}, their energy spectra are related as
\begin{eqnarray}\label{56}
\begin{array}{cccc}
E_l=E^{(1)}_l+E_0, & E^{(1)}_0=0, & \psi_l=\psi^{(1)}_l,& l=0,1,2,.., \\
\\
E^{(2)}_l=E^{(1)}_{l+1},\\
\\
\psi^{(2)}_l = [E^{(1)}_{l+1}]^{-\frac{1}{2}}A\psi^{(1)}_{l+1},\\
\\
\psi^{(1)}_{l+1} =
[E^{(2)}_{l}]^{-\frac{1}{2}}A^\dagger\psi^{(2)}_{l}.
\end{array}
\end{eqnarray}
Therefore, if the eigenvalues and eigenfunctions of $H_1$ were known,
one could immediately derive the spectrum of $H_2$. However the
above relations only give the relationship between the eigenvalues
and eigenfunctions of the two partner Hamiltonians. The condition of
exact solvability is known as the shape invariance property.
This condition means that the pair of SUSY partner potentials
$U_{1,2}(x)$ are similar in shape and differ only in the parameters
that appear in them \cite{15}
\begin{eqnarray}\label{57}
U_2(x;a_1)=U_2(x;a_2)+{\cal R}(a_1),
\end{eqnarray}
where $a_1$ is a set of parameters and $a_2$ is a function of $a_1$.
Then, the eigenvalues of $H_1$ are given by
\begin{eqnarray}\label{58}
E^{(1)}_l={\cal R}(a_1)+{\cal R}(a_2)+...+{\cal R}(a_l),
\end{eqnarray}
and the corresponding eigenfunctions are
\begin{eqnarray}\label{59}
\psi_l=\prod^{l}_{m=1}\frac{A^\dagger(x;a_m)}{\sqrt{E_m}}\psi_0(x;a_{l+1}).
\end{eqnarray}
The shape invariance condition (\ref{57}) can be rewritten in terms
of the factorization operators defined in equation (\ref{48})
\begin{eqnarray}\label{60}
A(x;a_1)A^\dagger(x;a_1) = A^\dagger(x;a_2) A(x;a_2)+{\cal R}(a_1),
\end{eqnarray}
where $a_2=f(a_1)$. Now we are ready to obtain spectra of $\Delta_l$
operator defined in (\ref{28}). For a given eigenspectrum of $E_l$,
we introduce the following factorization operators
\begin{eqnarray}\label{61}
\begin{array}{cc}
A_l=\frac{d}{dx}+l\tanh(x),\\
\\
A^\dagger_l=-\frac{d}{dx}+l\tanh(x),
\end{array}
\end{eqnarray}
the operator $\Delta_l$ can be factorized as
\begin{eqnarray}\label{62}
\begin{array}{cc}
 A^\dagger_l(x)A_l(x)\psi^{(1)}_n(x)=E^{(1)}_n\psi^{(1)}_n(x),\\
 \\
 A_l(x)A^\dagger_l(x)\psi^{(2)}_n(x)=E^{(2)}_n\psi^{(2)}_n(x).
 \end{array}
 \end{eqnarray}
 Therefore, for a given $l$, its first bounded excited state can be obtained
 from the ground state of $l-1$ and consequently the excited state $m$ of
 a given $l$, namely $\psi_{l,m}(x)$, can be written using (\ref{59}) as
 \begin{eqnarray}\label{63}
 \psi_{l,m}(x) = \sqrt{\frac{2(2m-1)!}{\Pi_{j=1}^mj(2l-j)}}\frac{1}{2^m(m-1)!}A^\dagger_l(x)
 A^\dagger_{l-1}(x)...A^\dagger_{m+1}(x)\frac{1}{\cosh^m(x)},
 \end{eqnarray}
 with eigenvalue $E_{l,m}=m(2l-m)$. Obviously its ground state with $E_{l,0}=0$
 is given by $\psi_{l,0}\propto \cosh^{-l}(x)$. Also its continuous spectrum
 consists of
 \begin{eqnarray}\label{64}
 \psi_{l,k}(x)=\frac{A^\dagger_{l}(x)}{\sqrt{k^2+l^2}}\frac{A^\dagger_{l-1}(x)}{\sqrt{k^2+(l-1)^2}}
 ...\frac{A^\dagger_{1}(x)}{\sqrt{k^2+1}}\frac{e^{ikx}}{\sqrt{2\pi}},
 \end{eqnarray}
 with eigenvalues $E_{l,k}=l^2+k^2$ and the following  normalization condition  \begin{eqnarray}\label{65}
 \int_{-\infty}^\infty \psi^*_{l,k}(x)\psi_{l,k'}(x)dx=\delta(k-k').
 \end{eqnarray}
 Therefore, using equations (\ref{40}), (\ref{41}), (\ref{63}) and (\ref{64})
 we find
 \begin{eqnarray}\label{66}
 G_{\Delta_l(0)}(x,y,\tau)=\frac{e^{-l^2\tau}}{2\sqrt{\pi \tau}}e^{-(x-y)^2/4\tau},
 \end{eqnarray}
 and
 \begin{eqnarray}\label{67}
 \begin{array}{cc}
 G_{\Delta_l}(x,y,\tau)=\sum_{m=1}^{l-1}\psi^*_{l,m}(x)\psi_{l,m}(y)e^{-m(2l-m)\tau}\\
 \\
 +\int_{-\infty}^\infty  \frac{dk}{2\pi}\frac{e^{-(l^2+k^2)\tau}}{\prod_{m=1}^l(k^2+m^2)}\left(\prod_{m=1}^lA^\dagger_m(x)e^{ikx}\right)^*
\left(\prod_{m=1}^lA^\dagger_m(y)e^{iky}\right).
\end{array}
\end{eqnarray}

In the case of closed Friedmann-Robertson-Walker universe we left with $l=2$ and then using
(\ref{42}) we have
\begin{eqnarray}\label{69}
\begin{array}{cc}
\xi_{\delta_2}(s)-\xi_{\delta_2(0)}(s)=3^{-s}-\frac{3}{\pi}\int_{-\infty}^\infty
\frac{dk}{(k^2+4)^{s+1}}=\\
\\
3^{-s}-\frac{3}{\sqrt{\pi}}2^{-(2s+1)}\frac{\Gamma(s+\frac{1}{2})}{\Gamma(s+1)}.
\end{array}
\end{eqnarray}

Consequently 
\begin{eqnarray}\label{71}
\begin{array}{cc}
\delta_1\varepsilon^k(s)=\frac{\hbar}{2}\mu^{2s+1}(\frac{mc^2}{\hbar})^{-2s}\left[\xi_{P\Delta_{l+h}}(s)-\xi_{\Delta_{(v)l+h}}(s)\right]_{l=2}=\\
\\
\frac{\hbar}{2}\mu^{2s+1}(\frac{mc^2}{\hbar})^{-2s}
\left(3^{-s}-\frac{3}{\sqrt{\pi}}2^{-(2s+1)}\frac{\Gamma(s+\frac{1}{2})}{\Gamma(s+1)}]\right).
\end{array}
\end{eqnarray}
Also we obtain
\begin{eqnarray}\label{72}
\begin{array}{cc}
\delta_2\varepsilon^k(s)= \lim_{L\rightarrow
\infty}\frac{\hbar}{2L}(\frac{mc^2}{\hbar})^{-2s}\mu^{2s+1}\frac{\Gamma(s+1)}{\Gamma(s)}
\xi_{\Delta_l+h(v)}(s+1)|_{l=2}\int_{-\frac{L}{2}}^\frac{L}{2}dx(-6\cosh^{-2}(x))\\
\\
=
-\frac{3\hbar}{\sqrt{\pi}}\mu^{2s+1}(\frac{mc^2}{\hbar})^{-2s}\frac{\Gamma(s+\frac{1}{2})}{\Gamma(s)}.
\end{array}
\end{eqnarray}
Finally we have
\begin{eqnarray}\label{73}
\begin{array}{ccc}
\lim_{s\rightarrow
-\frac{1}{2}}(\delta_1\varepsilon^k(s)+\delta_2\varepsilon^k(s))=\frac{\hbar}{2}\mu^{2s+1}(\frac{mc^2}{\hbar})^{-2s}\left(3^{-s}-\frac{3}{\sqrt{\pi}}\frac{\Gamma(s+\frac{1}{2})}{\Gamma(s+1)}-\frac{6}{\sqrt{\pi}}\frac{\Gamma(s+\frac{1}{2})}{\Gamma(s)}\right)\\
\\
=\frac{\hbar}{2}\frac{mc^2}{\hbar}\left(\sqrt{3}-\frac{3}{\pi}\right).
\end{array}
\end{eqnarray}
At last, we find the one-loop correction to the
kink energy as
\begin{eqnarray}\label{74}
\varepsilon^k=mc^{2}+\frac{mc^2}{2}(\sqrt{3}-\frac{3}{\sqrt{\pi}}).
\end{eqnarray}
Now, the renormalized gravitational constant $G_{one-loop}$ is obtained through
$G=\frac{\frac{3}{2}\pi c^4}{mc^2}$ as
\begin{eqnarray}\label{75}
G_{one-loop}=\frac{G}{1+\frac{1}{2}(\sqrt{3}-\frac{3}{\sqrt{\pi}})},
\end{eqnarray}
which indicates that the one-loop corrected Newtonian constant is smaller
than the original Newtonian constant.

\section{Conclusion}

The shape invariance property of the fluctuation operator is of particular
importance in order to find the exact one-loop quantum correction to the mass of kink solutions, namely the instanton solutions of classical field equations. The (1+1)-dimensional Sine-Gordon and $\phi^4$ field theories are some of these examples. This kind of potential occurs in different fields of physics particularly in quantum gravity and cosmology. In this paper,
we have shown that the system of closed Friedmann-Robertson-Walker universe is equivalent to the classical configuration space of the static field
$\Phi(x)$. Therefore, we expect kink-like solutions in the system of closed Friedmann-Robertson-Walker universe. The one-loop quantum corrections to the mass quantity of closed Friedmann-Robertson-Walker universe, namely $m=3\pi c^2/(2G)$, is therefore equivalent to the one-loop quantum corrections to the kink mass or ground state energy of the scalar field $\Phi(x)$. In fact,
what is physically meant by the one-loop quantum corrections to the ground state energy in FRW universe is nothing but the one-loop quantum corrections to the mass quantity $m=3\pi c^2/(2G)$ in FRW universe which plays the role of kink's mass or energy in the equivalent system of the classical static field $\Phi(x)$ configuration. We have therefore computed the quantum corrections to the ground state energy of the Kink-like solutions of closed Friedmann-Robertson-Walker universe. From mass-energy equivalence principle we know that any corrections to the energy is equivalent to the corresponding corrections to an equivalent mass quantity. In FRW cosmology this mass quantity coincides with the gravitational constant as $m=3\pi c^2/(2G)$. Therefore, it is conjectured that the one-loop quantum corrections to the ground state energy of the Kink-like solutions of closed Friedmann-Robertson-Walker universe leads to the renormalized gravitational constant which turns out to be smaller than the original gravitational constant. The calculated mass shift confirms the old result of Dashen {\it et al} in which they obtained extended objects which can be considered as prototypes for hadrons and studied their quantum corrections and renormalization \cite{16}.
The obtained correction may become more
viable and important whenever we study the quantum cosmology
of closed Friedmann-Robertson-Walker universe. For example, this correction
may have considerable impact in evaluation of the tunneling rate from "nothing" to a closed FRW universe with decaying cosmological term \cite{8'}. In \cite{8'},
the units was so taken that $8\pi G=1$. However, if we include $G$ explicitly
in the calculations we will find the classical action explicitly dependent on the gravitational constant. In fact, since the tunneling rate depends on the classical action which itself is dependent on the gravitational constant, then any renormalization on the gravitational constant will alter the tunneling rate. This is an interesting task which deserves to be further investigated.

\section*{Acknowledgment}
The authors would like to thank the anonymous referee whose useful comments
much improved the presentation of this paper. This work  has been supported by the Research office of Azarbaijan University of Tarbiat Moallem, Tabriz, Iran.

\end{document}